\documentclass[12pt,epsf,amssymb]{article}

\usepackage{graphicx}
\usepackage{amsfonts}
\usepackage{amsmath}
\usepackage[usenames]{color}
\usepackage{amssymb}
\usepackage[mathscr]{eucal} 
\usepackage{url}

\def\Z{\mathbb{Z}}
\def\Q{\mathbb{Q}}
\def\R{\mathbb{R}}
\def\C{\mathbb{C}}
\def\P{\mathbb{P}}

\begin{document}

\baselineskip 0.6cm
\newcommand{\vev}[1]{ \left\langle {#1} \right\rangle }
\newcommand{\bra}[1]{ \langle {#1} | }
\newcommand{\ket}[1]{ | {#1} \rangle }
\newcommand{\Dsl}{\mbox{\ooalign{\hfil/\hfil\crcr$D$}}}
\newcommand{\nequiv}{\mbox{\ooalign{\hfil/\hfil\crcr$\equiv$}}}
\newcommand{\nsupset}{\mbox{\ooalign{\hfil/\hfil\crcr$\supset$}}}
\newcommand{\nni}{\mbox{\ooalign{\hfil/\hfil\crcr$\ni$}}}
\newcommand{\nin}{\mbox{\ooalign{\hfil/\hfil\crcr$\in$}}}
\newcommand{\Slash}[1]{{\ooalign{\hfil/\hfil\crcr$#1$}}}
\newcommand{\EV}{ {\rm eV} }
\newcommand{\KEV}{ {\rm keV} }
\newcommand{\MEV}{ {\rm MeV} }
\newcommand{\GEV}{ {\rm GeV} }
\newcommand{\TEV}{ {\rm TeV} }

\def\diag{\mathop{\rm diag}\nolimits}
\def\tr{\mathop{\rm tr}}

\def\Spin{\mathop{\rm Spin}}
\def\SO{\mathop{\rm SO}}
\def\O{\mathop{\rm O}}
\def\SU{\mathop{\rm SU}}
\def\U{\mathop{\rm U}}
\def\Sp{\mathop{\rm Sp}}
\def\SL{\mathop{\rm SL}}

\def\change#1#2{{\color{blue}#1}{\color{red} [#2]}\color{black}\hbox{}}

\begin{titlepage}
 
\begin{flushright}
IPMU16-0126
\end{flushright}
 
\vskip 1cm
\begin{center}
 
 {\large \bf Vector-Like Pairs and Brill--Noether Theory}
 
\vskip 1.2cm
 

Taizan Watari
 
\vskip 0.4cm
 
%
   Kavli Institute for the Physics and Mathematics of the Universe, 
   University of Tokyo, Kashiwa-no-ha 5-1-5, 277-8583, Japan

\vskip 1.5cm
   
\abstract{How likely is it that there are particles in a vector-like pair of 
representations in low-energy spectrum, when neither symmetry nor anomaly 
consideration motivates their presence? We address this question in the context 
of supersymmetric and geometric phase compactification of F-theory and Heterotic 
dual. Quantisation of the number of generations (or net chiralities in more general 
term) is also discussed along the way. Self-dual nature of the fourth cohomology 
of Calabi--Yau fourfolds is essential for the latter issue, while we employ 
Brill--Noether theory to set upper bounds on the number $\ell$ of vector-like pairs 
of chiral multiplets in the ${\rm SU}(5)_{\rm GUT}$ $({\bf 5}+\bar{\bf 5})$ representations. 
For typical topological choices of geometry for F-theory compactification for SU(5) 
unification, the range of $0 \leq \ell \lesssim 4$ for perturbative unification 
is not in immediate conflict with what is already understood about F-theory 
compactification at this moment. } 
\end{center}
\end{titlepage}
 
 
\section{Introduction}

``{\it Who ordered that?}'' The Standard Model of particle physics contains 
three generations of quarks and leptons. Particle theorists have long been 
wondering what can be read out from the number of generations, $N_{\rm gen} = 3$.
If the Standard Model as a low-energy effective theory is obtained  
as a consequence of compactification of a high-energy theory in higher 
dimensional space-time, $N_{\rm gen}$ is often determined by index theorem (or an 
equivalent topological formula) on some internal geometry. Historically, 
it was first considered to be 
$\chi(Z; T^*Z) = \chi(Z)_{\rm top}/2$, the Euler characteristic of the cotangent 
bundle of a Calabi--Yau threefold $Z$, in a (2,2) compactification of 
Heterotic string theory \cite{Candelas}. Its generalisation in Heterotic 
string (0,2) compactifications is $\chi(Z; V)$, where $V$ is a vector bundle 
on $Z$. In Type IIB / F-theory language, $N_{\rm gen}$ is given 
by $\chi(\Sigma; K_{\Sigma}^{1/2} \otimes {\cal L}) = c_1({\cal L})$, where 
${\cal L}$ is a line bundle on a holomorphic curve $\Sigma$ in a complex threefold 
$M_{\rm int}$. 
In any one of those implementations, the fact that $N_{\rm gen}=3$ only means 
that one number characterising topology of compactification data happens 
to be 3.

Study of string phenomenology in the last three decades provides a dictionary 
of translation between the data of effective theory models and those for 
compactifications. An important question, then, is whether such a dictionary 
is useful.\footnote{Such a dictionary is important for those who ask 
whether string theory is able to reproduce the Standard Model at low-energy 
(if the answer is no, we should rule out string theory as the theory of 
quantum gravity!), even if the dictionary may not come with practical 
benefits (usefulness) in understanding the low-energy Lagrangian of this universe better.} 
The former group of data have direct connection with experiments, while 
we need to be lucky to have an experimental access to the latter in a near future; 
this means that the dictionary may not be testable. 
Compactification data may still provide correlations/constraints through 
the dictionary among various pieces of information in the effective theory
model data---that is the remaining hope. From this perspective, it is crucial 
which observable parameter constrains compactification data more. 
This letter shows, in section \ref{sec:Ngen}, that the value of $N_{\rm gen}$ brings 
virtually no constraint on the topology of the curve $\Sigma$, threefold $M_{\rm int}$ 
or $Z$;
this is due to the self-dual nature of the middle dimensional cohomology group of 
Calabi--Yau fourfolds, in F-theory language. This is a good news for those who seek 
for existence proof of appropriate compactifications, and a bad news for those who 
seek for profound meaning in $N_{\rm gen}=3$.  

In section \ref{sec:VectLike}, we focus on the number of matter fields in a vector-like 
pair of representations, as in the title of this article. It has often been adopted 
as a rule of game in bottom-up model building that vector-like pairs of matter fields 
are absent unless their mass terms are forbidden by some symmetry. Papers from 
string phenomenology community, on the other hand, often end up with such vector-like 
pairs in low-energy spectrum; difficulty of removing them from the spectrum is reflected 
the best in the heroic effort the U. Penn group had to undertake until they find a 
Heterotic compactification with just one pair of Higgs doublets. We will see, 
in section \ref{sec:VectLike}, that there is no reason to trust the bottom up 
principle based on the current understanding of F-theory/Heterotic string 
compactification; in the meanwhile, there is a good reason to believe 
(cf \cite{1408Andreas}) that generic vacua 
of F-theory compactification (and Heterotic dual) will predict smaller number 
of vector-like pairs than in papers (such as \cite{Penn-5, Hayashi-Flavor}) that have 
been written.\footnote{In this article, we are concerned about vector-like pairs in 
string compactification that are not associated in any way with symmetry or anomaly 
(and its flow). In compactifications that have an extra U(1) symmetry (which may be 
broken spontaneously or at non-perturbative level), low-energy spectrum tends to be 
richer, partially due to the 6D box anomaly cancellation of U(1) 
(cf \cite{Madrid-anomaly}). This article is concerned about more conservative set-ups, 
where there may or may not be an extra U(1) symmetry; matter parity is enough for 
SUSY phenomenology. } Brill--Noether theory 
sets upper bounds on the number of vector-like pairs $\ell$ for a given genus $g$ of 
a relevant curve $\Sigma$; given the typical range ${\cal O}(10)$--${\cal O}(100)$ for 
$g(\Sigma)$ for the matter fields in the ${\rm SU}(5)_{\rm GUT}$-$({\bf 5}+\bar{\bf 5})$ 
representations, the range of $0 \leq \ell \lesssim 4$ for perturbative unification 
are not in immediate conflict with most of internal geometry for F-theory / Heterotic 
string compactifications. 

Discussions in section \ref{sec:Ngen} and section \ref{sec:VectLike} are mutually 
almost independent. Despite many math jargons, logic of section \ref{sec:VectLike} 
will be simple enough for non-experts to follow. Observations in both sections 
will have been known to stringpheno experts already to some extent (e.g. section 7 
of \cite{1408Andreas}), but have not been written down as clearly and 
in simple terms as in this article, to the knowledge of the author. So, there will be a 
non-zero value in writing up an article like this. 

Language of supersymmetric and geometric phase F-theory compactification is used in 
most of discussions in this article. Heterotic string compactification on elliptic 
fibred Calabi--Yau threefolds is also covered by the same discussion, due to the 
Heterotic--F-theory duality. It is worth noting that large fraction of Calabi--Yau 
threefolds admit elliptic fibration \cite{CY-mostly-ell}.\footnote{M-theory 
compactification on $G_2$-holonomy manifolds is not discussed here, because the author 
is not a big fan of it. It is difficult to obtain realistic flavour pattern in SU(5) GUT 
in that framework \cite{TW-06}, and a solution to this problem has not been known so far.
If SU(5) unification is not used as a motivation, however, almost all kinds of string 
vacua (including IIA, IIB, Type I and those in non-geometric phase) will be just 
as interesting.}

\section{Quantisation of the Number of Generations}
\label{sec:Ngen}

{\large \bf Self-dual Lattice}

Let $X$ be a compact real $2n$-dimensional oriented manifold. 
Combination of the Poincare duality and the universal coefficient theorem implies that 
the middle dimensional homology group $[H_{n}(X; \Z)]_{\rm free}$ forms a self-dual 
lattice;\footnote{
This is not necessarily an even lattice. In the 2nd homology group 
of del Pezzo surfaces, for example, exceptional curves have odd 
self-intersection numbers. The 4th homology group of the sextic 
complex fourfold ($(6) \subset \P^5$) is not an even lattice either; 
the signature of this lattice is (1754, 852), where the difference 
$1754-852 = 902$ is not divisible by 8.} the intersection pairing matrix in  
\begin{equation}
 \left[ H_{n}(X; \Z) \right]_{\rm free} \times 
 \left[ H_{n}(X; \Z) \right]_{\rm free} \longrightarrow  \Z
\end{equation}
is symmetric and integer-valued, and its determinant is $\pm 1$.

There are several useful properties of self-dual lattices. 
Let $L$ be a self-dual lattice, $M$ a primitive non-degenerate sublattice of $L$, and  
$M' := [M^\perp \subset L]$ its orthogonal complement in $L$. 
The dual lattices of $M$ and $M'$ are denoted by $M^\vee$ and $(M')^\vee$, 
respectively. The intersection pairing of $L$ induces a homomorphism 
$L \longrightarrow M^\vee$. 
When $L$ is self-dual, this homomorphism is surjective, and the kernel is $M'$.
This homomorphism induces an isomorphism between $L/(M \oplus M')$ and a 
finite group $M^\vee/M$ (cf \cite{Nikulin}).

\vspace{5mm}

{\large \bf F-theory Applications}

\vspace{3mm}

{\bf Warming-up} 
We begin with the simplest example imaginable.
Consider using the sextic fourfold $X = (6) \subset \P^5$ for M-theory 
compactification. We have an effective theory of 2+1-dimensions then.

For a generic complex structure of $X$, algebraic two-cycles (real four-cycles) generate 
a rank-1 sublattice $M := \Z\vev{H^2|_X}$ of $L := H_4(X; \Z)$; the 
generator\footnote{An element $x$ of a lattice is a primitive element, if its 
self-intersection $(x,x)$ is not divisible by the square of an integer. 
Thus, in particular, $H^2|_X$ is a primitive element in $H^2 \in H_4(X; \Z)$.}
 is $H^2|_X$, where $H$ is the hyperplane divisor of $\P^5$, 
and $(H^2, H^2)=6$. Let $M' := [M^\perp \subset L]$ be the orthogonal complement of $M$. 
Since the dimension of the primary horizontal and primary vertical components of 
$H^{2,2}(X; \R)$---$h^{2,2}_H(X)$ and $h^{2,2}_V(X)$---add up to be $h^{2,2}(X)$ 
in this example, $M'\otimes \R \subset L \otimes \R$ corresponds to the primary 
horizontal component of $X = (6) \subset \P^4$. $M'$ must be a lattice of rank-$(b^4(X)-1)$ 
whose intersection form is given by a matrix with the determinant 6. 
Due to the property of self-dual lattices stated earlier, 
$L/(M \oplus M') \cong M^\vee/M \cong \Z_6$; 
we can choose $(1/6) \times H^2|_X$ mod $M$ as a generator of $M^\vee/M$.

When a fourform is restricted within a class 
\begin{equation}
 G = \left(\frac{15}{2} + n \right) H^2|_X 
     \in \frac{c_2(TX)}{2} + M \qquad {}^\forall n \in \Z , 
\end{equation}
it is guaranteed to be purely of (2,2) Hodge component for any complex 
structure of the sextic fourfold. Its integral over the algebraic 
cycle $H^2|_X$ can take a value in  
\begin{equation}
  \int_{H_2|_X} G = (H^2|_X, \; G) = 
     \left(\frac{15}{2} + n \right) \times 6 = 45 + 6n \in 3 + 6 \Z; 
\end{equation}
the value is quantised in units of 6, and cannot be 0, 1, 2, 4 or 5 modulo 6. 
When we allow the flux to be in \cite{Becker-Witten-quant}
\begin{equation}
G \in \frac{c_2(TX)}{2} + L,  
\end{equation}
however, the self-dual nature of the lattice $L = H_4(X; \Z)$ indicates 
that the integral $\int_{H^2|_X} G = (H^2|_X, \; G)$ can take any integer value. 
Such a flux $G$ is not purely of (2,2) Hodge component in an arbitrary 
complex structure of $X$, but the Gukov--Vafa--Witten superpotential drives 
the complex structure of $X$ to an F-term minimum, where the (1,3) and (3,1) 
Hodge components of the flux $G$ vanish (see also a comment later)\footnote{
Gravitino mass (the 
(4,0) and (0,4) Hodge components) does not vanish automatically, though. 
This is a well-known problem of supersymmetric compactification of string theory.
This article has nothing more to say about this. In this article, vanishing gravitino 
mass is not imposed, when we refer to supersymmetric compactifications.}

{\bf SU(5) GUT models}: Let us consider F-theory compactification on a fourfold $X_4$ 
so that there is a stack of 7-branes along a divisor $S$ in $B_3$. This means that 
there is an elliptic fibration $\pi: X_4 \longrightarrow B_3$, there is 
a section $\sigma: B_3 \longrightarrow X_4$, and $X_4$ has a locus of 
codimension-2 $A_4$ singularity in $\pi^{-1}(S)$. Let $\hat{X}_4$ be a 
non-singular Calabi--Yau fourfold obtained by resolving singularities 
of $X_4$ (see \cite{EY, Sakura-flux}  for conditions to impose on $\hat{X}_4$).

For concreteness of presentation, we choose the base threefold to be a $\P^1$-fibration 
over $\P^2$, 
\begin{equation}
 B_3 = \P\left[ {\cal O}_{\P^2} \oplus {\cal O}_{\P^2}(-n) \right], 
  \qquad -3 < n < 3, 
\label{eq:choice-B3}
\end{equation}
and the 7-brane locus $S$ to be the zero locus of the fibre of 
${\cal O}_{\P^2}$. $H^{1,1}(\hat{X}_4)$ is generated by $\sigma$, $S$, 
$H_{\P^2}$ and $E_{1,2,3,4}$, where $H_{\P^2}$ is the hyperplane divisor of 
$\P^2$, and $E_{1,2,3,4}$ are the Cartan divisors---the exceptional divisors 
emerging in the resolution of the $A_4$ singularity in $X_4$. The vertical 
component of $H^{2,2}(\hat{X}_4)$ is of 9 dimensions. We choose the 
following cycles as a set of generators of the vertical component of 
$H^{2,2}(\hat{X}_4)$: the first four are  
\begin{equation}
 \sigma \cdot S, \quad \sigma \cdot H_{\P^2}, \quad 
 S \cdot H_{\P^2}, \quad H_{\P^2} \cdot H_{\P^2},
\end{equation}
and the remaining five 
\begin{equation}
  E_i \cdot H_{\P^2} \quad (i=1, \cdots, 4) \quad {\rm and}  \qquad 
  E_2 \cdot E_4.
\end{equation}
These 9 cycles generate a rank-9 sublattice $M_{\rm vert}$ of a self-dual 
lattice $L = H_4(\hat{X}_4; \Z)$. The intersection form is given by 
\begin{equation}
  \left( \begin{array}{cccc}
   -n(3+n) & -(n+3) & n & 1 \\
    -(n+3) & 2 & 1 & 0 \\
    n & 1 & 0 & 0 \\
    1 & 0 & 0 & 0 \\
 \end{array} \right) 
 \oplus 
 \left( \begin{array}{cccc|c} 
    -2 & 1 & & & (3+n) \\
    1 & -2 & 1 & & -(3+n) \\
      & 1 & -2 & 1 & (3+n) \\
      &   & 1 & -2 & -(3+n) \\
    \hline
   (3+n) & - (3+n) & (3+n) & - (3+n) & -n(3+n) \\
\end{array} \right)
\end{equation}
in the basis of those 9 cycles; the determinant of this $9 \times 9$ matrix is 
${\rm discr}(M_{\rm vert})=(3+n)(18+n)$, which does not vanish in the range 
$-3 < n < 3$ of our interest. 

It is not obvious whether the lattice $M_{\rm vert}$ generated by the nine elements above 
is a primitive sublattice of $L$; since $L$ is not necessarily an even lattice, we have 
a limited set of tools to address this question. When it is not, however, we just have 
to replace the nine generators appropriately, so that $M_{\rm vert}$ becomes the primitive 
sublattice of $L$. Arguments in the following needs to be modified accordingly, but 
not in an essential way. ${\rm discr}(M_{\rm vert})$ may not be the same as $(3+n)(18+n)$ 
after the replacement, but the sublattice $M_{\rm vert}$ still remains non-degenerate.

Let $M'$ be the orthogonal complement, $[M^\perp \subset L]$, in the lattice $L$.
In the examples considered here, $M'$ corresponds to the horizontal components, 
$M_{\rm horz}$, because $M \otimes \Q= M_{\rm vert} \otimes \Q$ and the non-vertical 
non-horizontal component is empty \cite{1408Andreas}.  The quotient 
\begin{equation}
 L/(M \oplus M') \cong H_4(\hat{X}; \Z) / ( M_{\rm vert} \oplus M_{\rm horz}) 
\end{equation}
is a finite group isomorphic to $M^\vee/M = M_{\rm vert}^\vee/M_{\rm vert}$.

For a flux $G$ to preserve the SO(3,1) and SU(5) symmetry, it has to satisfy
all of \cite{DS}
\begin{equation}
  (G, \; x) = 0  \qquad {\rm for~} 
   x = \sigma \cdot S, \; \sigma \cdot H_{\P^2}, \; S \cdot H_{\P^2}, \; 
    H_{\P^2}^2, \; E_i \cdot H_{\P^2} \; (i=1,2,3,4).
 \label{eq:sym-cond}
\end{equation}
When we choose a fourform flux $G$ from $c_2(T\hat{X}_4)/2 + M$, 
the conditions above leave 
\begin{equation}
 G_{\rm FMW} = \lambda_{\rm FMW} \left(5 E_2 \cdot E_4 +
            (3+n)H_{\P^2} \cdot (2E_1 - E_2 + E_3 - 2E_4)\right), \qquad 
   \lambda_{\rm FMW} \in \frac{1}{2} + \Z, 
\label{eq:FMW-flux}
\end{equation}
as the only possible choice. This flux is always of pure (2,2) Hodge 
component for any complex structure of $\hat{X}_4$, and hence defines 
a supersymmetric vacuum. This is the flux constructed in \cite{FMW}; 
see \cite{CD, DW, Hayashi-NewAspects, Sakura-flux}. 
Within this class of choice of the fourform flux, the number of generations 
is quantised as follows \cite{Curio-DI}:
\begin{equation}
 N_{\rm gen} = (E_2 \cdot E_4, \; G) = \lambda_{\rm FMW} (3+n)(18+n); 
\end{equation}
although $\lambda_{\rm FMW}$ can change its value by $\pm 1$, $N_{\rm gen}$ cannot 
change by $\pm 1$. This would serve as a tight constraint in search of 
a geometry with ``right topology'' for the real world; 
the value of $|\lambda_{\rm FMW} (3+n)(18+n)|$ would never be as small as $3$ for 
the choice of $(B_3, S)$ we made here. 

In fact, we do not have to choose the flux from $c_2(T\hat{X}_4)/2 + M$. 
The condition of \cite{Becker-Witten-quant} does not rule out choice of flux 
from a broader class $c_2(T\hat{X}_4)/2 + L$.
Because of the self-dual nature of $L$, the homomorphism 
$L \longrightarrow M^\vee$ is surjective. This means that 
we can change the flux by $\Delta G \in L$ whose image in $M^\vee$ is 
anything one likes. In particular, there exists a change $\Delta G \in L$ 
so that $(\Delta G, \; x) = 0$ for all the eight generators 
in (\ref{eq:sym-cond}), while $N_{\rm gen}$ is changed by 
$(\Delta G, \; E_2 \cdot E_4) = \pm 1$. Therefore, the flux $G$ can be 
chosen within $c_2(T\hat{X}_4)/2 + L$ so that $N_{\rm gen}=3$, 
and the SO(3,1) and SU(5) symmetry is preserved.   Certainly such a flux 
is not purely of (2,2) Hodge component for generic complex structure of 
$\hat{X}_4$, but the complex structure of $\hat{X}_4$ is driven to 
an F-term minimum of the Gukov--Vafa--Witten superpotential, where the 
(1,3) + (3,1) Hodge component of the flux is absent automatically, and the 
moduli are stabilised (cautionary remark follows shortly, however).  

To put it from a slightly different perspective, the surjectivity 
of the homomorphism $L \longrightarrow M^\vee$ means that we can 
choose the $M^\vee \subset M \otimes \Q$ component of the 
flux in $L \otimes \Q$ arbitrarily, to suit the need from phenomenology 
(such as symmetry preservation and choosing $N_{\rm gen}$); this is, in effect, 
to relax the condition $\lambda_{\rm FMW} \in (1/2)+\Z$ and allow the overall 
coefficient (denoted $\lambda$ instead of $\lambda_{\rm FMW}$) to take any value 
in $[1/(3+n)(18+n)] \times \Z$. Once the $M^\vee$ component is chosen, then one 
can always find some element in $(M')^\vee$ so that their sum fits within 
$L \subset (M^\vee \oplus (M')^\vee)$. Depending upon phenomenological input, 
such as $N_{\rm gen}=3$, we may not be able to choose the flux so that the 
$(M')^\vee$ component vanishes, but that is an advantage rather than 
a problem, since complex structure moduli of $\hat{X}_4$ tend to be stabilised then. 

One can see that the $M^\vee$-component of the flux, 
(\ref{eq:FMW-flux}) with a relaxed quantisation in $\lambda$, satisfies 
the primitiveness condition $J \wedge G = (t_S S + t_{\P^2} H_{\P^2}) \cdot G = 0$, where 
$J$ is the K\"{a}hler form on $B_3$. This is enough to conclude that 
the primitiveness condition is satisfied, because the non-vertical 
component does not contribute to $J \wedge G$.

A cautionary remark is in order here. 
\label{pgrf:caution}
First, the $(M')^\vee = M_{\rm horz}^\vee$ component 
of the flux $G_{\rm horz}$ needs to be chosen so that $(G_{\rm horz})^2 > 0$, or otherwise 
there is no chance of finding a supersymmetric vacuum. This condition is not hard 
to satisfy, because we can change $G_{\rm horz}$ freely by $+M_{\rm horz}$ 
without changing the value of $N_{\rm gen}$ or breaking the SO(1,3) and SU(5) symmetry, 
and the lattice $M_{\rm horz}$ is not negative definite. An open question is, for a given 
$[G_{\rm horz}] \in M^\vee_{\rm horz}/M_{\rm horz}$, how one can find out whether there is a 
choice of Hodge structure of $\hat{X}_4$ so that there exists 
$G_{\rm horz} \in M^\vee_{\rm horz}$ with the vanishing negative component; note that a  
choice of Hodge structure introduces a decomposition of $M_{\rm hor} \otimes \R$ 
into $(2h^{4,0}+h^{2,2}_H)$-dimensional positive definite directions and 
$2h^{3,1}$-dimensional negative definite directions.\footnote{Even 
when such $G_{\rm horz} \in M^\vee_{\rm horz}$ and an appropriate Hodge structure 
is present, too large a positive value of $(G_{\rm horz})^2$ would violate the D3-tadpole 
condition. So, this is another physics condition to be imposed.} Due to the absence 
of a convenient Torelli theorem for general Calabi--Yau fourfolds, the author does not 
have a good idea how to address this problem. 

{\bf Generalisation}: The argument above can be used in set-ups where more phenomenological 
requirements are implemented. One can impose an extra U(1) symmetry (for spontaneous R-parity 
violation scenario instead of $\Z_2$ parity), and a flux for 
$\SU(5) \rightarrow {\rm SU}(3)_C \times {\rm SU}(2)_L \times {\rm U}(1)_Y$ symmetry 
breaking can be introduced in the non-vertical non-horizontal component of 
$H_4(\hat{X}_4)$ \cite{GUT-break}. One just has to take the lattice 
$M \subset L= H_4(\hat{X}_4; \Z)$ so it contains all the cycles relevant to symmetry 
(symmetry breaking) and the net chiralities of various matter representations in the 
low-energy spectrum. The self-dual nature of $H_4(\hat{X}_4;\Z)$ is the only essential 
ingredient in the argument above, and hence the same argument applies to more general 
cases.\footnote{The algebraic cycles $S$ to be used in $\chi = \int_S G$ to determine 
net chiralities need to be primitive elements of the primitive sublattice 
$M \subset L$, for the argument to apply.  
If some cycle $S$ were an integer multiple of another topological cycle, $m S'$ for 
some $m \in \Z$, then the net chirality on $S$ is always divisible by $m$, no matter 
how we choose a flux. The Madrid quiver \cite{Madrid}---fractional D3-branes at 
$\C^3/\Z_3$ singularity---is the best known example of that kind. The matter curve 
is effectively the canonical divisor of the vanishing cycle $\P^2$ at $\C^3/\Z_3$. 
$-3H$ is not primitive. } 

\vspace{5mm}

{\large \bf Heterotic Dual}

\vspace{3mm}

The same story should hold true, when the argument above in F-theory language 
is translated into the language of Heterotic string. $N_{\rm gen}$ can be chosen 
as we want it to be, by choosing the value of $\lambda_{\rm FMW}$ characterising the vector 
bundle for Heterotic compactification not necessarily in $(1/2) + \Z$. Supersymmetry 
can still be preserved, presumably by choosing the complex structure of a Calabi--Yau 
threefold $Z$ and vector bundle moduli appropriately and introducing a threeform flux and 
non-K\"{a}hlerity of the metric on $Z$. It is hard to verify this statement directly 
in Heterotic string language, but that must be true, if we believe that there is one-to-one 
dual correspondence (even at the level of flux compactification) between elliptic 
fibred Calabi--Yau threefold compactification of Heterotic string and elliptic fibred 
K3-fibred Calabi--Yau fourfold compactification of F-theory.\footnote{
The author does not make a bet on whether the same statement applies to Heterotic string 
compactification on Calabi--Yau's that do not admit an elliptic fibration morphism. }

\section{Number of Vector-Like Pair Multiplets}
\label{sec:VectLike}

We often encounter in supersymmetric string compactification with SU(5)$_{\rm GUT}$ 
unification that there are multiple pairs of chiral multiplets in the 
${\rm SU}(5)_{\rm GUT}$-$\bar{\bf 5}+{\bf 5}$ representations left in the low-energy 
spectrum and no perturbation in moduli can provide large masses to those vector-like 
multiplets. A good example is the one in \cite{Penn-5}, where the low-energy spectrum 
has $34+N'$ chiral multiplets in the ${\bf 5}$ representation and $34+N'+N_{\rm gen}$ 
of those in the $\bar{\bf 5}$ representation.\footnote{In the example studied 
in \cite{Penn-5}, there are $N_{\rm gen}=3$ chiral multiplets in the 
${\rm SU}(5)_{\rm GUT}$-${\bf 10}$ representation, while there is none 
in the $\overline{\bf 10}$ representation.} 
The $N' >0$ copies of chiral multiplets in the ${\bf 5}+\bar{\bf 5}$ representations 
have $\Delta W = \phi \cdot \bar{\bf 5} \cdot {\bf 5}$ coupling with moduli fields $\phi$, 
but 34 other vector-like pairs remain in the low-energy spectrum (at least without 
supersymmetry breaking) in the example studied in \cite{Penn-5}. It is likely that 
those 34 vector-like pairs have nothing to do with some symmetry in the 4D effective 
theory. 

Symmetry has been one of the most important guiding principles in bottom-up 
effective theory model building for more than three decades. It has often been 
assumed in model building papers that matter fields in a vector-like pair of 
representations are absent in low-energy spectrum, unless their mass terms 
are forbidden by some symmetry principle. Does the bottom-up guiding principle 
overlook something in string theory, or is there something yet to be understood 
in string phenomenology?

This guiding principle in bottom-up model building corresponds to the following 
statement in mathematics. Let us first note\footnote{translation for bottom-up model 
builders: roughly speaking, the {\it holomorphic curve} here is a real 2-dimensional 
submanifold within a real 6-dimensional internal space $M_{\rm int}$. It corresponds, 
in Type IIB language, to intersection of a 7-brane and another 7-brane, each one of 
which is wrapped on a 4-dimensional submanifold of $M_{\rm int}$ (and $\R^{3,1}$).  
The {\it line bundle} or {\it flux} here means gauge field configuration on the 
4-dimensional submanifold. 
Even in large fraction of Heterotic string compactifications, the number of 
the vector-like pairs $\ell$ can be discussed essentially with the same language, 
due to the duality between Heterotic string and F-theory. } 
that the number of 
${\rm SU}(5)_{\rm GUT}$-${\bf 5}$ and $\bar{\bf 5}$ chiral multiplets are given by 
\begin{align}
 h^0(\Sigma, {\cal O}(D)) \qquad {\rm and} \qquad h^1(\Sigma, {\cal O}(D)),
\end{align}
respectively, for some holomorphic curve $\Sigma$ and a line bundle ${\cal O}(D)$ on $\Sigma$,
quite often in supersymmetric and geometric phase compactifications of F-theory for 
SU(5) unification models \cite{FMW, Curio-DI, Penn-5, Munich, DW, Hayashi-NewAspects}. 
We assume that the flux (i.e., ${\cal O}(D)$) is chosen to realise 
the appropriate net chirality (cf discussion in section \ref{sec:Ngen})
\begin{align}
  \chi := h^0(\Sigma, {\cal O}(D)) - h^1(\Sigma, {\cal O}(D)),
\end{align}
which may be $-N_{\rm gen} = -3$ or 0, depending on whether or not the vector-like pairs 
are on the same complex curve as in the Standard Model $\bar{\bf 5}$'s. The number of 
extra vector-like pairs is 
\begin{align}
  \ell := h^0(\Sigma, {\cal O}(D)).
\end{align}
Now, it is known in mathematics \cite{BN} that $\ell = 0$ if 
\begin{itemize}
\item [(a)] the complex structure $\tau$ of $\Sigma$ is a generic element of the moduli space of the genus $g$ curve ${\cal M}_g$, and 
\item [(b)] the flux configuration ${\cal O}(D)$ is a generic element in ${\rm Pic}^{\chi+g-1}(\Sigma_g)$.
\end{itemize}
Thus, this general statement in math is in line with the bottom-up principle. The gap 
between the bottom-up guiding principle and the predictions of multiple vector-like pairs 
as in \cite{Penn-5, Hayashi-Flavor} must be due to non-genericity of the complex structure of 
the holomorphic curve, of the flux configuration, or of both, in the math moduli space 
${\cal M}_g$ and ${\rm Pic}^{\chi+g-1}(\Sigma_g)$. 

Most of papers for spectrum computation in F-theory or Heterotic string compactification 
so far employed the flux (\ref{eq:FMW-flux}) or something similar. With more general type 
of flux configuration (as discussed in section \ref{sec:Ngen}), however, more general 
elements of ${\cal O}(D) \in {\rm Pic}^{\chi+g-1}(\Sigma_g)$ can be realised than, 
for example, in \cite{Penn-5, Hayashi-Flavor}. Smaller number of vector-like pairs may 
be predicted in F-theory and elliptic fibred Heterotic string compactifications 
then (\cite{1408Andreas}).  

The question is how general $\tau \in {\cal M}_g$ and 
${\cal O}(D) \in {\rm Pic}^{\chi+g-1}(\Sigma)$ can be in such string compactifications. 
It is easy to see that the complex structure of the holomorphic curve $\Sigma$ for 
the ${\bf 5}+\bar{\bf 5}$ matter cannot be fully generic. Let us take the 
example (\ref{eq:choice-B3}) for illustration purpose. The genus $g$ of $\Sigma$ is 
given by \cite{AC, Hayashi-NewAspects}
\begin{align}
 2g-2 = (3n+24)(3n+21) - 2(3+n)(9+n) = 7n^2 + 111n + 450, 
\label{eq:genus-compt}
\end{align}
and the dimension of ${\cal M}_g$ is $3g-3$. On the other hand, the defining equation of 
the curve $\Sigma$ involves 
\begin{align}
  \left(\begin{array}{c} 5+n \\ 2 \end{array} \right)
+ \left(\begin{array}{c} 8+n \\ 2 \end{array} \right)
+ \left(\begin{array}{c} 11+n \\ 2 \end{array} \right)
+ \left(\begin{array}{c} 14+n \\ 2 \end{array} \right)
+ \left(\begin{array}{c} 21+n \\ 2 \end{array} \right)
- 9 = 
\frac{5n^2+113n+770}{2} 
 \label{eq:def-eq-dof}
\end{align}
complex parameters; the first five terms correspond to 
$h^0(\P^2; {\cal L})$ for line bundles ${\cal L} = {\cal O}(3+n)$, 
${\cal O}(6+n)$, ${\cal O}(9+n)$, ${\cal O}(12+n)$ and ${\cal O}(18+n)$; 
the last term accounts for the isometry of $\P^2$ and the overall scaling 
of the defining equation. The freedom (\ref{eq:def-eq-dof}) available for 
the complex structure of $\Sigma$ in F-theory compactification remains to be smaller 
than the $3g-3$ dimensions of the math moduli sapce ${\cal M}_g$, as long as 
$-3 \leq n$, which allows SU(5) GUT models.
The condition (a) necessary for the general math statement $\ell = 0$ (and 
absence of vector-like pairs) is not satisfied in string compactifications.\footnote{An intuitive (but not rigorous) alternative 
explanation is this. In Heterotic string, with a gauge field background in 
${\rm SU}(5)_{\rm str}$ (which breaks $E_8$ symmetry down to ${\rm SU}(5)_{\rm GUT} = 
[{\rm SU}(5)_{\rm str}^\perp \subset E_8]$), the $\bar{\bf 5}_{\rm GUT}$ matter fields are 
determined by the Dirac equation in the ${\bf 10} = \wedge^2 {\bf 5}_{\rm str}$ representation
 of ${\rm SU}(5)_{\rm str}$. Despite the 10 components participating in this Dirac equation, 
the structure group remains ${\rm SU}(5)_{\rm str}$, not SU(10).} We will also find more 
direct evidence for this in footnote \ref{fn:non-generic}.

To summarise, predictions of multiple vector-like pairs in string compactifications, 
such as those in \cite{Penn-5, Hayashi-Flavor}, do not have to be taken at face value, 
because only purely vertical flux was considered in those works; more generic choice 
(that involves horizontal components) would predict smaller number of vector-like pairs. 
But, the bottom-up guiding principle does not have to be trusted too seriously either,  
because the holomorphic curve $\Sigma$ for ${\rm SU}(5)_{\rm GUT}$-${\bf 5}+\bar{\bf 5}$ 
matter fields is not expected to have a generic complex structure. 

Brill--Noether theory \cite{BN}\footnote{
The phenomenon that the values of $\ell$ and $(\ell-\chi)$ jump up and down 
over the math moduli space ${\cal M}_g$ and ${\rm Pic}^{\chi + g-1}$ is a math translation 
of the coupling $\Delta W = z \cdot {\bf 5} \cdot \bar{\bf 5}$. The remaining question, 
which is partially discussed with (\ref{eq:def-eq-dof}) vs $(3g-3)$, is how much 
of the math moduli space is covered by the physical moduli space (fields) of 
compactification. 
In other words, it is to study $z(\phi, G)$, where $\phi$ denotes physical moduli 
and $G$ the flux.} tells us a little more than the general math statement quoted above. 
Let $\Sigma$ be a genus $g$ curve and ${\cal O}(D)$ a line bundle on $\Sigma$ whose 
degree is $d = \chi+g-1$. First of all, 
\begin{align}
  \ell = 0 \qquad {\rm if~} d < 0.
\end{align}
When $0 \leq d \leq g-1$, there are soft upper bound and hard upper bound. 
Clifford's theorem provides the hard upper bound, 
\begin{align}
  \ell \leq \frac{d}{2}+1 = \frac{\chi + g + 1}{2}, 
  \label{eq:BN-1}
\end{align}
which holds for any complex structure of smooth curve $\Sigma$. When the complex 
structure of $\Sigma$ is not non-generic, there is a stronger upper bound,\footnote{
\label{fn:non-generic}
This upper bound is not always satisfied (hence this is a soft upper bound), when the 
complex structure of $\Sigma$ is somewhat special. A good example is found in \cite{Hayashi-Flavor}. There, a flux is chosen as in 
(\ref{eq:FMW-flux}), including the quantisation condition on $\lambda_{\rm FMW}$, so that 
$\chi = - N_{\rm gen}= -17$. In addition to this net chirality in the 
${\rm SU}(5)_{\rm GUT}$-${\bf 5}+\bar{\bf 5}$ sector, non-removable $\ell = 11$ vector-like 
pairs are predicted in that example. In this case, $g = 174$, and hence $d = 156$. 
The hard upper bound $\ell \leq d/2+1 = 79$ is satisfied, but the stronger upper bound 
for $\Sigma$ with a generic complex structure, $\ell \leq 7.15$, is not satisfied. So, 
this computation is a direct evidence that the curve $\Sigma$ for the ${\bf 5}+\bar{\bf 5}$ 
matter in F-theory does have a special complex structure (even after choosing the complex 
structure of $\hat{X}_4$ completely generic). The dimension counting argument 
using (\ref{eq:def-eq-dof}) is not the only evidence for the non-genericity of 
$\tau \in {\cal M}_g$. It will be possible to carry out similar study for the 
examples in \cite{Timo}. } 
\begin{align}
  \ell \leq \frac{\chi+\sqrt{\chi^2+4g}}{2},
  \label{eq:BN-2}
\end{align}
because the Brill--Noether number $\rho := g - \ell(\ell - \chi)$ becomes negative 
for $\ell$ beyond this upper bound. Due to the Serre duality, it is enough to focus 
on the cases with $d \leq g-1$. 

In the case of ${\rm SU}(5)_{\rm GUT}$-$\overline{\bf 10}+{\bf 10}$ matter fields, 
string compactification often ends up with $g \leq -\chi = N_{\rm gen} = 3$ 
(though not always), and hence the $d<0$ case applies. 
The vector like pair of $\overline{\bf 10}+{\bf 10}$ is absent then. In the case of 
${\rm SU}(5)_{\rm GUT}$-${\bf 5}+\bar{\bf 5}$ matter fields, however, $g$ often takes a 
much larger value (as in the example (\ref{eq:genus-compt})), and hence the $\ell = 0$ 
result does not apply. Typical values of $g$ listed in Table 1 and 2 of \cite{Hayashi-Flavor} are in the range of 
${\cal O}(10)$--${\cal O}(100)$. For such large values of $g$, $d = \chi + g-1$ is close to 
$g-1$ for $\chi = - N_{\rm gen} = -3$ or $\chi = 0$. The upper bounds (\ref{eq:BN-1}, \ref{eq:BN-2}) for those $g$ and $d$ have no conflict with vector-like pairs in the range of 
$0 \leq \ell \lesssim 4$ for perturbative gauge coupling unification.\footnote{
The $H^{2,1}$ moduli of F-theory compactification (and also presumably their Heterotic dual)
do not receive large supersymmetric mass terms from the Gukov--Vafa--Witten superpotential, and are likely to change ${\cal O}(D) = K_{\Sigma}^{1/2}\otimes {\cal L}$ in ${\rm Pic}^{\chi+g-1}(\Sigma_g)$.
So they are good candidates of a singlet field $S$ that have a coupling 
$\Delta W = S \cdot {\bf 5} \cdot \bar{\bf 5}$; some of the $H^{3,1}$ moduli may also 
remain unstabilized supersymmetrically (i.e., in the low-energy spectrum) and play the 
same role. There is nothing new in that observation, but there will be some value 
to leave such a footnote in this article as a reminder. } 

It requires much more dedicated study to go beyond. One could try to characterise what 
the physically realised subspace---one with the dimension given in (\ref{eq:def-eq-dof})---in ${\cal M}_g$ would be like, or to work out the image 
of not necessarily purely vertical fluxes mapped into ${\rm Pic}^{\chi+g-1}(\Sigma)$; the 
cautionary remark in page~\pageref{pgrf:caution} also needs to be taken care of along 
the way. They are 
way too beyond the scope of this article, however. It is also worth studying how discussion 
in this article needs to be modified, when spontaneous R-parity violation scenario is at work
(where an off-diagonal 4D scalar field breaking a U(1) symmetry to absorb a non-zero 
Fayet--Iliopoulos parameter (cf section 5 of \cite{FMW} and \cite{stability-Sharpe, 
Penn-parabolic, FI-05-06, RparityViol-w, stability-wall-Het, offdiag-F})).

\subsection*{Acknowledgements}

I am grateful to T. Yanagida for encouraging me to address the issue of 
section \ref{sec:VectLike} for nearly a decade. Relevance of Brill--Noether theory 
to the issue was explained to me by R. Donagi at the workshop 
``Elliptic Fibrations and F-theory'' at Kavli IPMU, 2010.
This work is benefited from collaboration with A. Braun and Y. Kimura, and 
obviously, inspired by \cite{Strumia} and references therein. 
I am also grateful to Aspen Centre for Physics (Summer programme 2015) and Harvard 
theory group for hospitality, where a part of this work was done. This work is 
supported in part by WPI Initiative, Brain Circulation Programme and a Grant-in-Aid 
on Innovative Areas 2303, MEXT, Japan. 


\begin{thebibliography}{99}
%
\bibitem{Candelas}
%
P.~Candelas, G.~T.~Horowitz, A.~Strominger and E.~Witten,
  ``Vacuum Configurations for Superstrings,''
  Nucl.\ Phys.\ B {\bf 258} (1985) 46.
%
%
\bibitem{Penn-5}
%
R.~Donagi, Y.~H.~He, B.~A.~Ovrut and R.~Reinbacher,
  ``The Particle spectrum of heterotic compactifications,''
  JHEP {\bf 0412} (2004) 054
  [hep-th/0405014].
%
%
\bibitem{Hayashi-Flavor}
%
Section 3.2 of 
%
H.~Hayashi, T.~Kawano, Y.~Tsuchiya and T.~Watari,
  ``Flavor Structure in F-theory Compactifications,''
  JHEP {\bf 1008} (2010) 036
  [arXiv:0910.2762 [hep-th]].
%
Its section 2 is also remotely relevant to this article. 
%
\bibitem{Madrid-anomaly}
%
G.~Aldazabal, D.~Badagnani, L.~E.~Ibanez and A.~M.~Uranga,
  ``Tadpole versus anomaly cancellation in D = 4, D = 6 compact IIB orientifolds,''
  JHEP {\bf 9906} (1999) 031
  [hep-th/9904071].
%
\bibitem{1408Andreas}
%
%
A.~P.~Braun and T.~Watari,
  ``The Vertical, the Horizontal and the Rest: anatomy of the middle cohomology of Calabi-Yau fourfolds and F-theory applications,''
  JHEP {\bf 1501} (2015) 047
  [arXiv:1408.6167 [hep-th]].
%
%
\bibitem{CY-mostly-ell}
%
J.~Gray, A.~S.~Haupt and A.~Lukas,
  ``Topological Invariants and Fibration Structure of Complete Intersection Calabi-Yau Four-Folds,''
  JHEP {\bf 1409} (2014) 093
  [arXiv:1405.2073 [hep-th]].
%
\bibitem{TW-06}
%
R.~Tatar and T.~Watari,
  ``Proton decay, Yukawa couplings and underlying gauge symmetry in string theory,''
  Nucl.\ Phys.\ B {\bf 747} (2006) 212
  [hep-th/0602238].
%
\bibitem{Nikulin}
%
V.~Nikulin, ``Integral symmetric bilinear forms and some of their applications,'' 
Math. USSR Izv. {\bf 14} (1980) 103--167. 
[Russian original Izv. Akad. Nauk SSSR Ser. Mat. {\bf 43} (1979) 111--177.] 
%
\bibitem{Becker-Witten-quant}
%
K.~Becker and M.~Becker,
  ``M theory on eight manifolds,''
  Nucl.\ Phys.\ B {\bf 477} (1996) 155
  [hep-th/9605053].
%
E.~Witten,
  ``On flux quantization in M theory and the effective action,''
  J.\ Geom.\ Phys.\  {\bf 22} (1997) 1
  [hep-th/9609122].
%
\bibitem{EY}
%
M.~Esole and S.~T.~Yau,
  ``Small resolutions of SU(5)-models in F-theory,''
  Adv.\ Theor.\ Math.\ Phys.\  {\bf 17} (2013) 1195
  [arXiv:1107.0733 [hep-th]].
%
\bibitem{DS}
%
 K.~Dasgupta, G.~Rajesh and S.~Sethi,
  ``M theory, orientifolds and G - flux,''
  JHEP {\bf 9908} (1999) 023
  [hep-th/9908088].
%
\bibitem{FMW}
%
R.~Friedman, J.~Morgan and E.~Witten,
  ``Vector bundles and F theory,''
  Commun.\ Math.\ Phys.\  {\bf 187} (1997) 679
  [hep-th/9701162].
%
\bibitem{CD}
%
G.~Curio and R.~Y.~Donagi,
  ``Moduli in N=1 heterotic / F theory duality,''
  Nucl.\ Phys.\ B {\bf 518} (1998) 603
  [hep-th/9801057].
%
%
\bibitem{DW}
%
R.~Donagi and M.~Wijnholt,
  ``Model Building with F-Theory,''
  Adv.\ Theor.\ Math.\ Phys.\  {\bf 15} (2011) 1237
  [arXiv:0802.2969 [hep-th]].
%
\bibitem{Hayashi-NewAspects}
%
H.~Hayashi, R.~Tatar, Y.~Toda, T.~Watari and M.~Yamazaki,
  ``New Aspects of Heterotic--F Theory Duality,''
  Nucl.\ Phys.\ B {\bf 806} (2009) 224
  [arXiv:0805.1057 [hep-th]].
%
\bibitem{Sakura-flux}
%
J.~Marsano, N.~Saulina and S.~Schafer-Nameki,
  ``On G-flux, M5 instantons, and U(1)s in F-theory,''
  arXiv:1107.1718 [hep-th].
%
\bibitem{Curio-DI}
%
G.~Curio,
  ``Chiral matter and transitions in heterotic string models,''
  Phys.\ Lett.\ B {\bf 435} (1998) 39
  [hep-th/9803224].
%
D.~E.~Diaconescu and G.~Ionesei,
  ``Spectral covers, charged matter and bundle cohomology,''
  JHEP {\bf 9812} (1998) 001
  [hep-th/9811129].
%
\bibitem{Madrid}
%
G.~Aldazabal, L.~E.~Ibanez, F.~Quevedo and A.~M.~Uranga,
  ``D-branes at singularities: A Bottom up approach to the string embedding of the standard model,''
  JHEP {\bf 0008} (2000) 002
  [hep-th/0005067].
%
\bibitem{GUT-break}
%
M.~Buican, D.~Malyshev, D.~R.~Morrison, H.~Verlinde and M.~Wijnholt,
  ``D-branes at Singularities, Compactification, and Hypercharge,''
  JHEP {\bf 0701} (2007) 107
  [hep-th/0610007].
%
C.~Beasley, J.~J.~Heckman and C.~Vafa,
  ``GUTs and Exceptional Branes in F-theory - II: Experimental Predictions,''
  JHEP {\bf 0901} (2009) 059
  [arXiv:0806.0102 [hep-th]].
%
\bibitem{Munich}
%
R.~Blumenhagen, S.~Moster, R.~Reinbacher and T.~Weigand,
  ``Massless Spectra of Three Generation U(N) Heterotic String Vacua,''
  JHEP {\bf 0705} (2007) 041
  [hep-th/0612039].
%
\bibitem{AC}
%
B.~Andreas and G.~Curio,
  ``On discrete twist and four flux in N=1 heterotic / F theory compactifications,''
  Adv.\ Theor.\ Math.\ Phys.\  {\bf 3} (1999) 1325
  [hep-th/9908193].
%
\bibitem{BN}
%
E.~Arbarello, M.~Cornalba, P.A.~Griffiths, J.~Harris, 
``Geometry of Algebraic Curves,'' vol. I, Springer--Verlag, 1985. \\
%
A quick review is found in section 2 of 
%
S.~Mukai, ``Vector bundles and Brill-Noether theory,'' 
MSRI Publ. {\bf 28}, 145--158, {\it Cambridge Univ. Press}, 1995.
%

\bibitem{Timo}
%
M.~Bies, C.~Mayrhofer, C.~Pehle and T.~Weigand,
  ``Chow groups, Deligne cohomology and massless matter in F-theory,''
  arXiv:1402.5144 [hep-th].
%
\bibitem{Strumia}
%
A.~Strumia,
  ``Interpreting the 750 GeV digamma excess: a review,''
  arXiv:1605.09401 [hep-ph].
%

\bibitem{stability-Sharpe}
%
E.~R.~Sharpe,
  ``Kahler cone substructure,''
  Adv.\ Theor.\ Math.\ Phys.\  {\bf 2} (1999) 1441
  [hep-th/9810064].
%
R.~Donagi, S.~Katz and E.~Sharpe,
  ``Spectra of D-branes with higgs vevs,''
  Adv.\ Theor.\ Math.\ Phys.\  {\bf 8} (2004) no.5,  813
  [hep-th/0309270].
%
\bibitem{Penn-parabolic}
%
V.~Braun, Y.~H.~He, B.~A.~Ovrut and T.~Pantev,
  ``The Exact MSSM spectrum from string theory,''
  JHEP {\bf 0605} (2006) 043
  [hep-th/0512177].
%
V.~Bouchard and R.~Donagi,
  ``An SU(5) heterotic standard model,''
  Phys.\ Lett.\ B {\bf 633} (2006) 783
  [hep-th/0512149].
%
\bibitem{FI-05-06}
%
R.~Blumenhagen, G.~Honecker and T.~Weigand,
  ``Loop-corrected compactifications of the heterotic string with line bundles,''
  JHEP {\bf 0506} (2005) 020
  [hep-th/0504232].
%
Also \cite{TW-06, Munich}.
%
\bibitem{RparityViol-w}
%
%
M.~Kuriyama, H.~Nakajima and T.~Watari,
  ``Theoretical Framework for R-parity Violation,''
  Phys.\ Rev.\ D {\bf 79} (2009) 075002
  [arXiv:0802.2584 [hep-ph]].
%
%
\bibitem{stability-wall-Het}
%
L.~B.~Anderson, J.~Gray, A.~Lukas and B.~Ovrut,
  ``The Edge Of Supersymmetry: Stability Walls in Heterotic Theory,''
  Phys.\ Lett.\ B {\bf 677} (2009) 190
  [arXiv:0903.5088 [hep-th]].
%
L.~B.~Anderson, J.~Gray, A.~Lukas and B.~Ovrut,
  ``Stability Walls in Heterotic Theories,''
  JHEP {\bf 0909} (2009) 026
  [arXiv:0905.1748 [hep-th]].
%
L.~B.~Anderson, J.~Gray and B.~Ovrut,
  JHEP {\bf 1005} (2010) 086
  [arXiv:1001.2317 [hep-th]].
%
L.~B.~Anderson, J.~Gray and B.~A.~Ovrut,
  Fortsch.\ Phys.\  {\bf 59} (2011) 327
  [arXiv:1012.3179 [hep-th]].
%
\bibitem{offdiag-F}
%
 R.~Tatar, Y.~Tsuchiya and T.~Watari,
  ``Right-handed Neutrinos in F-theory Compactifications,''
  Nucl.\ Phys.\ B {\bf 823} (2009) 1
  [arXiv:0905.2289 [hep-th]].
%
S.~Cecotti, C.~Cordova, J.~J.~Heckman and C.~Vafa,
  ``T-Branes and Monodromy,''
  JHEP {\bf 1107} (2011) 030
  [arXiv:1010.5780 [hep-th]].
%
 R.~Donagi and M.~Wijnholt,
  ``Gluing Branes II: Flavour Physics and String Duality,''
  JHEP {\bf 1305} (2013) 092
  [arXiv:1112.4854 [hep-th]].
%
\end{thebibliography}
\end{document}